\documentclass[twocolumn,showpacs,preprintnumbers,amsmath,amssymb]{revtex4}
\usepackage{graphicx}

\draft

\begin{document}


\title{ Hydrogen in Ag-doped ZnO: theoretical calculations}
\author{
        H.Y. He, J. Hu and B.C. Pan
}
\affiliation{
 Hefei National Laboratory for Physical Sciences at Microscale and Department of Physics, \\
 University of Science and Technology of China,\\
 Hefei, Anhui 230026, People's Republic of China \\
}


\begin{abstract}

Based on density functional theory calculations, we systematically investigate the behaviors of a H atom 
in Ag-doped ZnO, involving the preference sites, diffusion behaviors, the electronic 
structures and vibrational properties. We find that a H atom can migrate to the doped Ag to form 
a Ag-H complex by overcoming energy barriers of 0.3 - 1.0 eV. 
The lowest-energy site for H location is the bond center of a Ag-O in the basal plane. Moreover, H can migrate 
between this site and its equivalent sites with energy cost of less than 0.5 eV. 
In contrast, dissociation of such a Ag-H complex needs energy of about 1.1 - 1.3 eV. This implies that 
the Ag-H complexes can commonly exist in the Ag-doped ZnO, which have a negative effect on the desirable p-type 
carrier concentrations of Ag-doped ZnO.
In addition, based on the frozen phonon calculation, the vibrational properties of ZnO with a Ag-H complex 
are predicted. Some new vibrational modes associated with the Ag-H complex present 
in the vibrational spectrum of the system.

\end{abstract}


\maketitle

\section {Introduction}

ZnO is a promising material for short wavelength optoelectronic device, due to its various attractive properties, 
such as optical \cite{Look1, Lee1, Kang}, photoelectric \cite{Wenas} and piezoelectric \cite{Mitsuyu} properties.
However, the difficulty in fabricating p-type ZnO restricts the 
application of ZnO in future \cite{Zhang}. 
Therefore, many efforts have been made to achieve p-type ZnO, with using many techniques and dopants.  
Among various dopants, either group I elements (Li, Na, and K) or group IB 
elements (Cu, Ag, and Au) \cite{Zunger, Zhang2, Wei} are 
good candidates for substitution of Zn.
However, with smaller ionic radii, group I elements prefer to 
occupy the interstitial sites rather than substitutional sites, acting as donors \cite{Look}. 
In contrast, group IB elements have suitable size for substitution of Zn. So far, the Ag doped p-type ZnO
has been successfully achieved in the experiment \cite{Kang2}.

On the other hand, the role of H in ZnO was paid much attention, 
because H is usually to be unintentionally-doped in as-grown ZnO bulk, and it always acts as a ``hole-killer" 
in ZnO \cite{Walle1, Cox, Hofmann}. Previous studies revealed that the activation energy for H diffusion in 
pure ZnO was about 0.17 - 0.50 eV \cite{Ip, Nickel,Wardle}, which indicated that a H atom could
migrate in ZnO easily at low temperature. In ZnO with p-type dopants such as N, Cu and Li, H can be easily trapped by 
the dopants to form the impurity complexes. 
For instance, in N doped ZnO, H preferably located
at the antibond site of N-Zn bond to form a N-H complex \cite{Zhang1}, and the N-H vibrational modes were observed 
by Raman spectroscopy in experiment \cite{Kumar}. While in the case of Li doped ZnO, the preferred site for H 
was the bond center of a Li-O bond parallel to the $c$ axis. Such a Li-H complex was studied 
both experimentally and theoretically \cite{Lavrov3,Halliburton,Wardle2,Shi,Martin}. 
Except for the Li-H and N-H complexes, the complexes of H with transition metals such 
as Cu, Fe and Ni, were also studied \cite{Wardle1,Lavrov2}, in which presence of H significantly 
reduced the effect of the spin polarization.   
Up to date, behavior of H in Ag doped p-type ZnO has not been reported yet.

In this paper, based on density functional theory calculations, we find that a H atom can be trapped by 
the doped Ag atom in ZnO to form a Ag-H complex, and the lowest-energy 
site (ground state site) is the bond center of Ag-O in the basal plane.
In addition, the vibrational features associated with the Ag-H complex are also discussed in this paper. 

\section {Computational details}

Our calculations are based on the density functional theory implemented in the SIESTA program \cite{Siesta}.
Double- $\zeta$ basis plus polarization (DZP) sets\cite{Soler} are used for all the concerned atoms, and 
spin polarization is taken into account in our calculations.
 A 72-atom supercell consisting of $3\times3\times2$ primitive unit cells is employed, in which 
a Zn atom is substituted by a Ag atom, and a H atom is located at some typical sites around the doped Ag respectively. 
The lattice constants for each case are optimized and periodic boundary condition is applied.
The Brillouin zone is sampled with a set of $k$-point grids ($3\times3\times2$) according
to the Monkhorst-Pack scheme \cite{Monkhorst}. The configurations are fully relaxed with using 
the conjugate gradient method until the Hellman-Feynman force on each atom is less than 0.02 eV/{\AA}.

With these settings, we optimize the geometry of the perfect wurtzite ZnO ($w$-ZnO) both with the local 
density approximation (LDA) and with the Perdew-Burke-Ernzerhos generalized gradient approximation (GGA-PBE). 
The obtained optimal crystallographic parameters of $w$-ZnO from LDA calculations are $a$= 3.278 {\AA}, 
$c$=5.237 {\AA}, and $u$=0.382, which are slightly larger than those ( $a$ =3.25 {\AA}, $c$=5.207 {\AA}, and $u$=0.382) 
measured from the experiment \cite{Albertsson} at room temperature. 
It is noted that the previous experiment demonstrated that the lattice constants of ZnO decreased slightly when
the temperature of the system increased \cite{Yan}. Therefore, 
our calculated parameters are consistent with the experimental values at low temperature. 
In contrast, the lattice constants evaluated at the level of GGA are larger than those at the level of LDA.
On the other hand, the formation enthalpy (-3.43 eV/ZnO) of $w$-ZnO
obtained from the LDA calculation does also match that (-3.61 eV/ZnO) derived from the experiments.
Therefore, we perform our calculations at the level of LDA.
In addition, we test the convergence of the super cell size by calculating the
      defect formation energy of the $Ag_Zn $ in the supercells of $2\times2\times2$, $3\times2\times2$ and 
$3\times3\times3$ respectively, in which the lattice constants are optimized in each case. Here the defect formation 
energy is defined as the difference 
of the total energies between the defect case and the corresponding perfect case. We find that 
 the difference of the defect formation energies between $2\times2\times2$ and $3\times2\times2$ supercells is
as large as 0.5 eV, whereas for the supercells of $3\times3\times2$ and $3\times3\times3$, the difference of the 
defect formation energies is about 0.02 eV only.
This shows that the interaction of the doped Ag with its images is very weak in supercell $3\times3\times2$, 
and thus the influence of the images of Ag can be neglected. In addition, the obtained bond lengths as well as 
the electronic structures in $3\times3\times2$ are almost the same as that in $3\times3\times2$ supercell. 
So the results achieved from the $3\times3\times2$ supercell are reliable.

The frozen phonon approximation \cite{Yin} is employed to explore the vibrational properties 
of the typical systems with Ag-H complexes.
In our calculations, the atoms in each of the systems are displaced one by one from their equilibrium positions 
along three Cartesian directions and the reverse directions, with an amplitude of 0.04 Bohr.
By the numerical derivatives for the displacements of each atom, the force constants are obtained, which are used 
to build up the dynamical matrix of this system. By solving the dynamical equation
                                                                                                                             
\begin{equation}
    \omega^{2}M_{i}u_{i,\alpha} = \sum _{j,\beta} C_{i,\alpha;j,\beta}u_{j,\beta},
\end{equation}
                                                                                                                             
the vibrational frequencies $\omega$ and the corresponding eigenmodes $u_{i,\alpha}$ are yielded, where
the $M_{i}$ is the mass of $ith$ atom, $C_{i,\alpha;j,\beta}$ is the force constant, and $\alpha$ ($\beta$) means
the direction of x, y or z.

\section {Diffusion of a H atom in Ag-doped ZnO}

Firstly, we replace a Zn atom in the perfect ZnO supercell by a Ag atom, then optimize the system 
with the considerations above. 
Then a H atom is located around some typical O atoms,
which are notated as $O_I$, $O_{II}$, $O_{III}$, $O_{IV}$ and $O_{V}$ respectively, as shown in Fig.1. These O atoms 
are near the doped Ag within the fifth neighbour.
For the structure of ZnO bulk, there are two kinds of inequivalent $Zn-O$ bonds which is either parallel to $c$ axis or
within the basal plane respectively. Therefore, for the Ag-doped ZnO bulk, there are four 
inequivalent sites available for the location of H around each concerned O atom: 
bond center (BC) sites and antibond (AB) sites of $Zn(Ag)-O$ bonds parallel to the $c$ axis 
and those of $Zn(Ag)-O$ bonds in the basal plane.
From the calculations, we find that some of these sites are unstable for H, where the located H 
atom spontaneously moves to other places during full relaxation. For example, when a H atom is located at 
either AB site of $Zn-O_{II}$ bond parallel to the $c$ axis or BC site of $Zn-O_{II}$ bond in the basal plane, 
it spontaneously migrates to a BC site or AB site of $Ag-O_{II}$ bond during geometry optimization. 
For convenience, some stable or metastable sites for H are marked with characters of $A_{n}$ (n=1 to 3) for AB sites 
and $B_{n}$ (n=1 to 7) for BC sites in Fig.1. The coordinates of these sites 
and the doped Ag are listed in Table I.

\begin{table}
  \centering
  \caption{The optimized coordinates of H at the stable (metastable) sites near the doped Ag.
The coordinates of the doped Ag is also listed. The positions for H refer to Fig.1.
 The lattice vectors of the supercell are
$a$=(9.85, 0.0, 0.0), $b$=(-4.93, 8.53, 0.0), and $c$=(0.0, 0.0, 10.65).
The unit of the length is in angstrom.}
\begin{tabular}{c|c|c|c|c|c|c|c|c|c|c|c}
\hline
    & A1 & A2 & A3  & B1 & B2 & B3  & B4 & B5  & B6 & B7 & Ag   \\
\hline
   x           &1.71&1.64&-1.13 &1.65&0.63&-0.58 &1.07&-0.03&2.68&1.03&1.64   \\
   y           &2.87&2.84& 4.65 &2.86&3.43& 4.13 &3.14& 5.55&3.42&5.96&2.85   \\
   z           &8.80&3.37& 4.18 &6.62&4.91& 3.86 &1.92& 6.62&4.92&4.57&5.32   \\
 \hline
\end{tabular}
                                                                                                                             
\end{table}

Among these sites, site $B2$ in Fig.1 is the ground state site for H location, which is similar to 
that of H in Cu doped ZnO \cite{Hu}. Meanwhile, site $A3$ 
is another preference site for H, where the energy of the system is only 0.03 eV higher than that at site $B2$.
Nevertheless, the obtained energies for the other metastable sites are higher than that at site $B2$ by 
less than 1.0 eV.

\begin{figure}
\includegraphics[width=0.45\textwidth]{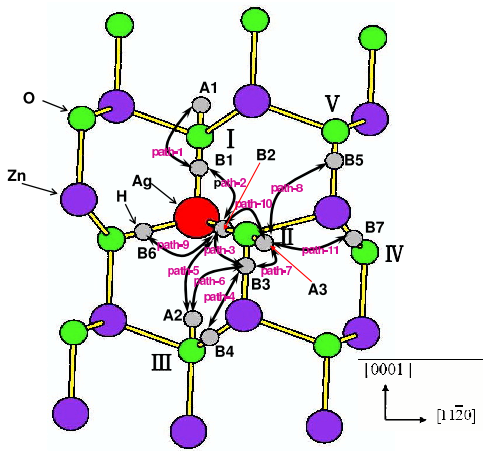}
\caption{(Color online) Schematic showing of the concerned diffusion paths of H
around the doped Ag in ZnO,
indicated by path-1 to path-11.
As labeled, green, purple, red and gray balls represent O, Zn, Ag and H atoms respectively.
Some O atoms near the Ag atom are marked with Romanic numbers. The stable and metastable sites
for H location in the diffusion paths are marked with symbols of $A_{n}$ (n=1 to 3) and $B_{n}$ (n=1 to 7).}\label{Fig.1}
\end{figure}

Since the ground state and metastable sites for H are obtained, 
we now study the diffusion behaviors of H near 
the doped Ag. Eleven diffusion paths are considered, which are marked in Fig.1.
By using the climbing image nudged elastic band (CI-NEB) scheme \cite{Jonsson1, Jonsson2},
the energies for the images in each diffusion path are calculated. Taking the lowest energy of the system 
(with H at site $B2$) as 
a reference, the energy profiles for each concerned diffusion path are plotted in Fig.2. 
As shown in Fig.2 (a), once a H atom arrives at the site of either $A1$ or $B1$ near $O_I$, 
it can migrate to the ground state site ($B2$), by overcoming energy barriers of less than 1.0 eV. 

In contrast, when a H atom reaches site $B4$ near $O_{III}$, it can diffuse to site $B2$ via site $B3$ (path-4, path-3), 
with energy cost of less than 0.5 eV. For a H atom at site $A2$, it prefers to migrate towards site $B2$ 
directly through path-5, rather than towards site $A3$ via site $B3$ (path-6, path-7), 
because the energy barrier in the former case is about 0.4 eV, while that in the latter case is about 0.7 eV, 
which can be seen in Fig.2 (b). 
Meanwhile, if a H atom reaches near $O_{V}$ (at site $B5$) or $O_{IV}$ (site $B7$), it can reach site $A3$ 
through path-8 or path-11, with overcoming a low barrier of about 0.4 or 0.3 eV, as shown in Fig.2 (b,c).
Interestingly, the energy barriers for H diffusing between the equivalent sites of $B2$ and $B6$ 
are less than 0.5 eV (Fig.2 (c), path-9), which indicates that a H atom can move "freely" between 
$B2$ and its equivalent sites near the doped Ag at room temperature. 
Such local motion of H near the doped Ag was also found in the Cu-doped ZnO \cite{Hu}. Moreover, H can diffuse between 
the low energy sites of $B2$ and $A3$ through two pathways. One is from site $B2$ 
to site $A3$ directly via path-10 with energy barriers of less than
0.9 eV, and the other is to site $B3$ through path-3 firstly, then to site $A3$ through path-7 
by overcoming the energy barriers of no more than 0.7 eV.
On the other hand, one can find from the Fig.2 that once a H atom locates at either site $B2$ or $A3$,
the dissociation of such a Ag-H complex needs the energy cost as high as about 1.1 - 1.3 eV.

\begin{figure}
\includegraphics[width=0.40\textwidth]{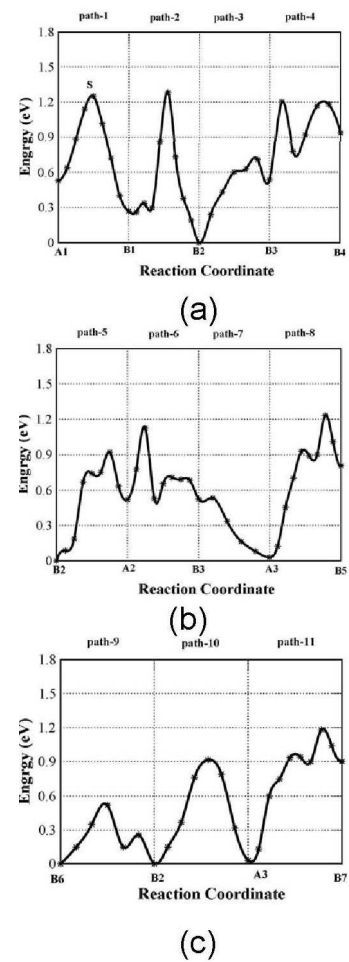} 
\caption{Energy profiles calculated with CI-NEB for the diffusion paths 
of H in Ag-doped ZnO: (a) from path-1 to path-4, (b) from path-5 to path-8, 
and (c) from path-9 to path-11.
The calculated relative energies for the images in each path are indicated with
stars.
The diffusion paths and reaction coordinates shown in this figure refer to Fig.1.}\label{Fig.2}
\end{figure}

From above, we can conclude that a H atom can diffuse to the doped Ag to form a Ag-H complex 
in ZnO through many paths, and the lowest diffusion barrier is 0.3 eV only. 
In contrast, the dissociation of a Ag-H complex requires energy 
costs of more than 1.1 eV. This indicates that the Ag-H complex may easily exist in the Ag-doped ZnO. 
We stress that such stable complexes of Ag-H in ZnO play a negative role in p-type ZnO with doped Ag.

To go further, we pay our attention to the influence of a Ag-H complex on electronic structures of the system. 
The calculated total density of states (TDOS) for the systems with a H atom at site $B2$ and $A3$ 
are shown in Fig.3 a and b respectively. 
For comparison, the TDOS for the Ag monodoped ZnO and the perfect ZnO with the same size are shown in Fig.3 (c). 
Obviously shown in Fig.3 (a and b), some occupied defect states appear near Fermi level 
for the cases of H at site $B2$ and site $A3$. 
By analysis of the local density of states (LDOS), we find that these defect states mainly come from 
the contribution of the doped Ag atom and its neighbouring O atoms, and little from H 
and its bonded O atom, as displayed in Fig.3 (d and e). Furthermore, the projected density of 
states (PDOS) analysis reveals that these states are mainly contributed from 3d orbitals of the Ag atom 
and 2p orbitals of the O atoms. 

\begin{figure}
\includegraphics[width=0.35\textwidth,angle=270]{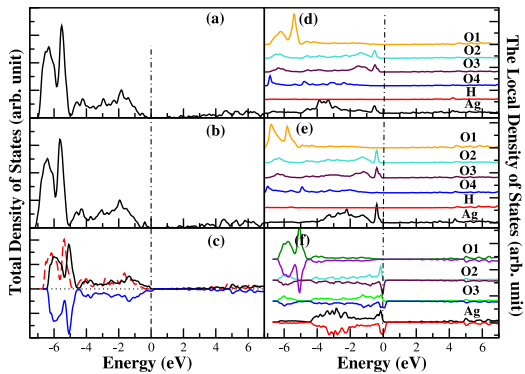}
\caption{(Color online) The total density of states (TDOS) of the system for (a) the Ag-H complex with H at site $B2$,
(b) the Ag-H complex with H at site $A3$, (c) ZnO with and without Ag doping,
and the corresponding local density of states (LDOS) (d), (e) and (f).
The dash line in (c) stands for the TDOS of the perfect ZnO, and the solid lines in (c) stand for
Ag-doped ZnO with majority spin and minority spin. O1 refers to the O atom being far away from the the doped Ag;
O2 and O3 stand for the O atoms in Ag-O parallel and perpendicular
to the $c$ axis respectively, and O4 for the O atom in H-O bond. The dot-dash lines indicate
the Fermi levels, which are shifted to be zero.}\label{Fig.3}
\end{figure}

\section {vibrational properties}

Vibration is a fundamental property of a system, which is closely coupled with the structure of the system.
For such a Ag-H complex in ZnO, some new vibrational features should appear in the vibrational spectrum. 
With the frozen phonon approximation, we explore the vibrational properties of the ZnO containing a Ag-H complex.
As a reference, we firstly calculate the total vibrational density of states (TVDOS) for the perfect ZnO, which is 
plotted in Fig.4 (b). The high frequency range for the perfect ZnO is below 570 $cm^{-1}$, and a silent region exists between
270 and 400 $cm^{-1}$. This is consistent with the previous report \cite{Hewot}.

\begin{figure}
\includegraphics[width=0.35\textwidth,angle=270]{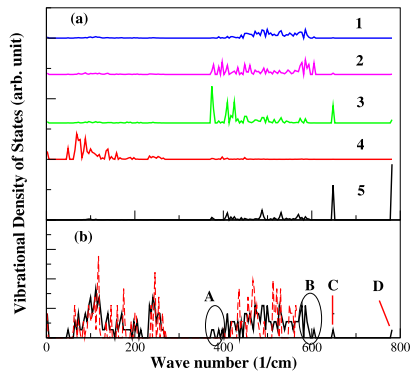}
\caption{(Color online) (a) The local vibrational density of states (LVDOS) and
(b) total vibrational density of
states (TVDOS) for the Ag-H complex with H at site $B2$, where frequency range is below 800 cm$^{-1}$.
The stretch modes of H-O are not included.
1,2,3,4 and 5 in (a) represent the averaged LVDOS for the O atoms far from
Ag, the O atoms around Ag, the O atom bonding with H, Ag and H respectively. In
(b),
the red dash line stands for the TVDOS of the perfect ZnO, and the solid line for that of the complex system with Ag-H. The
states marked with the characters indicate the new states induced by the Ag-H complex.}\label{Fig.4}
\end{figure}

For the system containing a Ag-H complex with H at site $B2$, 
a typical stretching mode of H-O bond is found to be at 3275 $cm^{-1}$, 
and no state between 782 and 3275 $cm^{-1}$. 
The calculated TVDOS in the region below 782 $cm^{-1}$ is shown in Fig.4 (b). From Fig.4 (b), we find that
some new vibrational states, which are marked
with $A$, $B$, $C$ and $D$ respectively, appear in the silent frequency region of the perfect ZnO.

The local vibrational density of states (LVDOS) analysis reveals that the states at 
lower frequency region of TVDOS (below 270 $cm^{-1}$) 
for the perfect ZnO comes from the contribution of the Zn atoms, and those at higher frequency 
region (above 400 $cm^{-1}$) from the O atoms. 
This remains in the TVDOS for ZnO containing a Ag-H complex.
Furthermore, we find that the vibrational states marked with characters of $A$ (below 400 $cm^{-1}$) and 
$B$ (above 570 $cm^{-1}$) in Fig.4 (b) are 
mainly from the contribution of four nearest neighbouring O atoms around Ag, and little from the O atoms 
far from Ag, which can be seen obviously in Fig.4 (a). 
The presence of these modes are ascribed to the local structural distortion induced by the Ag-H complex.
By examining the configuration of the system, we find that the Ag-O bonds both parallel and 
perpendicular to $c$ axis are all enlarged by about 0.2 {\AA}. Consequently, the Zn-O bonds in the vicinity 
are shortened or enlarged within 0.04 {\AA}, and the bond lengths of the next neighbouring 
Zn-O are altered slightly. Such structural changes result in 
presence of these new states near $A$ and $B$ in Fig.4 (b).      
Meanwhile, the vibrational states marked with $C$ (at about 646 $cm^{-1}$) and $D$ (782 $cm^{-1}$) are attributed 
to the bending modes of the H-O bond. In addition, the states contributed 
from the doped Ag mainly emerge at the low frequency region, as shown in Fig.4 (a).

For the case of a H atom at site $A3$, we also calculate the TVDOS of the system, and its main features 
are similar to those of H at site $B2$. The stretching mode of the H-O bond locates at the frequency of 3287 $cm^{-1}$,
and the bending modes of the H-O bond are at 768 and 879 $cm^{-1}$ respectively. 
Similarly, some new states associated with the neighbouring 
O atoms around the doped Ag emerge near the frequency regions of below 400 and above 570 $cm^{-1}$ respectively. 
Especially, a mode corresponding to displacement of H and its bonded O atom in the same direction is found at the frequency 
of 273 $cm^{-1}$. These typical vibrational modes may drip a hint to detect the presence of such a Ag-H complex in experiment.

\section {Summary}

In summary, the site preferences and diffusion behaviors of a H atom in Ag-doped ZnO are investigated based 
on density functional theory calculations. 
We find that the lowest-energy site for H in the Ag-doped ZnO is bond-centered sites
of Ag-O in the basal plane. H can migrate to these sites through some 
typical diffusion paths with energy barriers of 0.3 - 1.0 eV, most of which are less than 0.7 eV. 
H can diffuse between its ground state sites with the energy cost of less than 0.5 eV. 
In contrast, releasing of this H atom from the doped Ag requires energy of about 1.1 - 1.3 eV, which indicates 
that the Ag-H complex may commonly exist in Ag-doped ZnO.
By calculating the vibrational properties of ZnO with a Ag-H complex,
we find some new vibrational modes in the silent frequency region of perfect ZnO, which are attributed 
to the distortion of the O atoms nearby induced by the Ag-H complex.

\section {Acknowledgments}
This work is supported by the University of Science and Technology of China, the Chinese academy of
Science, National Science Foundation of China (Grant Nos. NSFC10574115 and NSFC50721091).
B. C. Pan thanks the support of National Basic Research Program of China (2006CB922000).
This indicates that the Ag-H complexes can commonly exist in ZnO.
The HP-LHPC of USTC is acknowledged for computational support.

\widetext
\end{document}